\documentclass[twocolumn,showpacs,preprintnumbers,amsmath,amssymb,floatfix,prb,superscriptaddress,hyperref]{revtex4}
\usepackage{xcolor,graphicx}
\usepackage{bm}
\usepackage{textcomp}

\begin{document}

\title{Rashba-Dresselhaus spin-splitting in the bulk ferroelectric oxide BiAlO$_3$}
\author{Luiz Gustavo Davanse da Silveira}
\affiliation{
Consiglio Nazionale delle Ricerche (CNR-SPIN), UOS L'Aquila, Sede di lavoro c/o University ``G. D'Annunzio'' Chieti-Pescara,
66100 Chieti, Italy} 

\author{Paolo Barone}
\affiliation{
Consiglio Nazionale delle Ricerche (CNR-SPIN), 67100 L'Aquila, Italy}
\affiliation{
Graphene Labs, Istituto Italiano di Tecnologia, via Morego 30, 16163 Genova, Italy
}

\author{Silvia Picozzi}
\affiliation{
Consiglio Nazionale delle Ricerche (CNR-SPIN), UOS L'Aquila, Sede di lavoro c/o Univ. ``G. D'Annunzio'' Chieti-Pescara,
66100 Chieti, Italy} 
\date{\today}

\begin{abstract}
It has been recently suggested that the coexistence of ferroelectricity and Rashba-like spin-splitting effects due to spin-orbit coupling in a single material may allow for a non-volatile electric control of spin degrees of freedom. In the present work, we compared the structural and ferroelectric properties of tetragonal and rhombohedral phases of ferroelectric  BiAlO$_3$ by means of density-functional calculations. In both phases, we carefully investigated Rashba and Dresselhaus effects, giving rise to spin-splitting  in their bulk electronic structure, particularly near the conduction band minimum, supplementing our first-principles results with an effective $\bm k\cdot\bm p$ model analysis. The full reversal of the spin texture with ferroelectric polarization switching was also predicted. BiAlO$_3$ can therefore be considered as the first known oxide to exhibit a coexistence of ferroelectricity and Rashba-Dresselhaus effects. 
\end{abstract}


\maketitle

\section{Introduction}

Generally considered a ``weak'' interaction, the spin-orbit coupling (SOC), a relativistic interaction arising from electrons movement in the nuclear electric field, has proven to be a key ingredient of new and exotic phenomena in solid state physics. 
In particular, in solids lacking an inversion center, the gradient of the crystalline electric potential will not vanish, since $V(r)\neq V(-r)$, resulting in an electric field. This field, coupled to  atomic SOC, leads to a spin-momentum coupling that lifts Kramers' degeneracy and spin-splits the electronic bands at wave vectors $\bm k$, which are not time-reversal invariant, even in the absence of magnetic fields. 

Dresselhaus \cite{PhysRev.100.580} was the first to demonstrate that in acentric nonpolar crystals (such as zincblende), SOC produces a spin-splitting proportional to $k^3$. On the other hand, in crystals with polar axis, such as wurtzite, also terms linear in $k$ are allowed, as shown by Rashba. \cite{Rashba1960} Considering a surface subjected to a normal electric field $E=E_z z$, Vas'ko \cite{Vasko1979} and Bychkov and Rashba \cite{Bychkov1979} proposed that the Rashba-SOC can be described by
\begin{equation}\label{H_Rashba}
H_R=\alpha_R\,(\bm\sigma\times \bm k)\cdot \hat{z} =\,\alpha_R\, (\sigma_x k_y - \sigma_y k_x),
\end{equation}   
where the so-called Rashba parameter $\alpha_R$ (proportional to $\lambda E_z$, $\lambda$ being the spin-orbit constant) represents the strength of the Rashba effect. While Eq. \ref{H_Rashba} is strictly correct only for plane-wave eigenstates as, e. g., for a two-dimensional electron gas,  \cite{Manchon2015} generally speaking the form of the spin-momentum coupling in bulk materials is determined by the symmetry properties of the wave functions in reciprocal space. Spin splitting linear in $k$ can be also realized in acentric but nonpolar structures, provided that they belong to gyrotropic point groups such as $D_{2d}$, in which case the spin-momentum coupling is given by the linear Dresselhaus SOC:\cite{winkler_book,ganichev_rev}
\begin{eqnarray}\label{H_Dresselhaus}
H_D=\alpha_D\, (\sigma_x k_x-\sigma_y k_y).
\end{eqnarray}
Solution of the free-electron Hamiltonian including the Rashba term, Eq. (\ref{H_Rashba}), yields split spin-polarized states (labeled $+$ and $-$) with energies:
\begin{equation}\label{E_split}
E_\pm(k)=\frac{\hbar^2k^2}{2m^*}\pm\alpha_R k.
\end{equation}
The Rashba momentum $k_R=m^*\alpha_R/\hbar^2$ quantifies the mutual shift of the split bands, with $E_R=\alpha_R^2m^*/(2\hbar^2)$ being the energy of the split band minimum. By neglecting higher-order terms in the nearly free-electron approximation, the Rashba parameter can be related to $E_R$ and $k_R$ by $\alpha_R = 2E_R/k_R$. Notice that the linear Dresselhaus term would lead to the same energy splitting; however, the specific symmetries of the Rashba and Dresselhaus wave functions give rise to different distributions of momentum-dependent spin orientations.\cite{winkler_book}

The research on Rashba effect has so far mostly focused on material surfaces and interfaces, where inversion symmetry is structurally broken. For example, Rashba spin-splitting has been observed at metallic surfaces,  \cite{PhysRevLett.77.3419,PhysRevLett.93.046403,PhysRevLett.98.186807} ultrathin metal films, \cite{PhysRevLett.101.196805,PhysRevLett.101.107604} and semiconductor heterostructures. \cite{PhysRevLett.78.1335} A major breakthrough was provided by BiTeI, the first known non-centrosymmetric semiconductor showing huge Rashba-like spin-splitting in its bulk (not surface) band-structure. \cite{Ishizaka2011,PhysRevLett.109.096803}

An interesting class of bulk materials lacking inversion symmetry is represented by ferroelectrics, i.e., materials displaying, below a certain critical temperature, a long-range dipolar order with a permanent ferroelectric (FE) polarization switchable by an electric field. Recently, the occurrence of the Rashba effect in the FE GeTe has been theoretically predicted \cite{Domenico2013} and experimentally confirmed.\cite{morgenstern,krempasky} GeTe is the prototype of a new class of multifunctional materials called ferroelectric Rashba semiconductors (FERSC), \cite{FERSC2014} which integrates different subfields: ferroelectricity, Rashba effect, and semiconductor spintronics. One of the most interesting features of FERSC is the link between the spin texture of split bands and ferroelectric polarization, implying a full reversal of the spin orientations when the latter is switched. GeTe itself, however, has several pitfalls from the experimental point of view. In fact, whereas ferroelectric displacements have been clearly observed with many experimental techniques, \cite{GeTe1987} GeTe shows a high tendency to form Ge vacancies. \cite{Kobolov2003,Giussani2012} This in turn leads to a $p$-degenerate semiconducting behavior, calling into question the possibility to switch the ferroelectric state in such a ``conducting material'', therefore hindering the control of spin-texture via an electric field (it is worth noticing that FE switching has been, however, recently reported in epitaxial GeTe films).\cite{calarco_switching} The search for new FERSC materials is hence essential to achieve better properties and performances. Apart from the possibilities for applications, studying materials that exhibit both ferroelectricity and Rashba effect is also interesting from a basic science standpoint, since the correlations between both phenomena are not yet completely understood.

In the present work, we focus on a ferroelectric material, BiAlO$_3$ (BAO). In 2005, a theoretical study predicted BAO to have a polar perovskite structure with space group \textit{R3c}. \cite{Baettig2005} This was later confirmed in samples synthesized on high temperature and high pressure conditions. \cite{Belik2006} The calculated polarization value\cite{Baettig2005} was of 75.6 $\mu C/cm^2$; however, reported values for measurements done in polycrystalline samples\cite{Zylberberg2007,Mangalam2008435} are 9.5  and 12 $\mu C/cm^2$. To the best of our knowledge, there is only one report about thin-film growth of BAO. \cite{Son2008} The thin-film sample was reported to have a tetragonal symmetry and polarization of 29 $\mu C/cm^2$ at room temperature. Different crystal symmetries and polarization values of bulk and thin films of BAO partly resemble the situation observed in BiFeO$_3$ (BFO). \cite{NatMater6,Catalan2009}

The manuscript is organized as follows: after reporting technicalities in Sec. \ref{sec:tech}, we compare the structural and ferroelectric properties of tetragonal and rhombohedral phases of BAO, calculated within density-functional theory (Sec. \ref{sec:structure}). 
Although some of the DFT results (for example,  ferroelectric polarization and displacements in rhombohedral BAO) were already reported in the literature by other authors, we repeat them here, as their detailed discussion is preliminary to the main focus of the paper: a   careful investigation of SOC-induced spin-splitting effects in the BAO electronic structure, both from first-principles and within a $\bm k \cdot\bm p$ model analysis (reported in Sec. \ref{sec:elec}). Finally, we draw our conclusions in Sec. \ref{sec:conclusion}.    

\section{Computational methods and structural details}\label{sec:tech}

The simulations were performed using the Vienna Ab-initio Simulation Package (VASP) \cite{PhysRevB.54.11169} within the density functional theory using the supplied PAW pseudopotentials \cite{PhysRevB.50.17953,PhysRevB.59.1758} and Perdew-Burke-Ernzerhof generalized gradient approximation (PBE-GGA). \cite{PhysRevLett.77.3865} The potential for Bi included the semicore \textit{d} states in the valence. Rhombohedral and tetragonal crystalline structures of BAO were examined. For the rhombohedral case (space group \textit{R3c}), the experimental lattice parameters and atomic positions \cite{Belik2006} were used as a starting point and a relaxation of the internal atomic positions was performed using a $6\times6\times6$ k-point mesh containing the $\Gamma$ point. The hexagonal setting for the unit cell was also considered, since it allows for a more direct and intuitive interpretation of the results. The unit cell in hexagonal setting is the same as in Refs. \onlinecite{Baettig2005,Belik2006,Zylberberg2007}. Due to the lack of published crystallographic data regarding the tetragonal phase of BAO, a prototypical centrosymmetric tetragonal structure (space group \textit{P4mm}) was constructed using the reported lattice parameters. \cite{Son2008} Then the atomic positions were relaxed using a $6\times6\times6$ k-point Monkhorst-Pack grid \cite{PhysRevB.13.5188} in order to obtain the ferroelectric structure. The band structure, density of states (DOS), and spin texture were calculated with SOC included. For the DOS calculation a $16\times16\times16$ k-point mesh was used for the tetragonal structure and a $11\times11\times11$ grid for the rhombohedral (for the DOS calculation the rhombohedral setting was used). Ferroelectric polarization has been evaluated in the framework of Berry-phase theory of polarization. \cite{PhysRevB.47.1651,RevModPhys.66.899} A cutoff energy of 550 $eV$ was employed in all calculations. For the structural optimization the change in total energy between two ionic relaxation steps was required to be smaller than $10^{-5}$ $eV$. The atomic positions were optimized without including SOC; however, test calculations showed no significant changes when considering SOC. For a better description of the excited state  properties, we also performed some benchmarks within  the accurate non local hybrid-functionals according to the Heyd-Scuseria-Erzenhof (HSE) approach. \cite{HSE2003}

\section{Structural and ferroelectric properties of rhombohedral and tetragonal phases}\label{sec:structure}

The structural data used in this work are presented in Tables \ref{tab:lattice} and \ref{tab:wyck}. Our calculations showed that the total energy per formula unit of the tetragonal phase is about 80 $meV$ higher than the rhombohedral energy. This difference is very similar to the one calculated between BFO rhombohedral and tetragonal phases, \cite{PhysRevB.83.094105} possibly implying that tetragonal BAO could be stabilized under strain (similarly to the situation of tetragonal BFO).\cite{PhysRevLett.95.257601,Ricinschi2006,PhysRevLett.102.217603,Kim2014} Furthermore, it has been  theoretically demonstrated that BFO can form a large variety of metastable structural phases, due to, in part, Bi's ability to form different coordination complexes with the neighboring oxygens. \cite{PhysRevB.83.094105} BAO possesses, in principle, the same potential to have a rich structural phase diagram as BFO, although such investigation is beyond the scope of the present work.

\begin{table}
\caption{Lattice parameters and calculated polarization for tetragonal and rhombohedral BiAlO$_3$ phases.}\label{tab:lattice}
\begin{tabular}{ccccc}
\hline
Space group & $a$ (\AA{}) & $c$ (\AA{})  & $V$ (\AA{}$^3$) & P ($\mu C/cm^2$) \\ 
\hline
$P4mm$ & 3.80 & 3.92 & 57.33 & 90.46 \\ 
$R3c$ & 5.38 & 13.39 & 335.16 & 79.38 \\ 
\hline
\end{tabular}
\end{table} 

\begin{table}
\caption{Relaxed atomic positions for tetragonal and rhombohedral BiAlO$_3$ phases.}\label{tab:wyck}
\begin{tabular}{cccccc}
\hline
Space group & Site & Wyckoff position & x & y & z \\ 
\hline
P4mm & Bi & 1a & 0 & 0 & 0 \\ 
 & Al & 1b & 0.5 & 0.5 & 0.4247 \\ 
 & O & 1b & 0.5 & 0.5 & 0.8899 \\ 
 & O  & 2c & 0.5 & 0 & 0.3650 \\ 
\hline
R3c & Bi & 6a & 0 & 0 & 0.9936 \\ 
 & Al & 6a & 0 & 0 & 0.2211 \\ 
 & O & 18b & 0.5493 & 0.0097 & 0.9606 \\
\hline 
\end{tabular}
\end{table}

For the rhombohedral phase, we calculate a change in ferroelectric polarization relative to the centrosymmetric structure of 79.38 $\mu C/cm^2$ along the [0001] direction in the hexagonal setting ([111] in the rhombohedral setting),consistent with a previously published study. \cite{Baettig2005} A larger ferroelectric polarization of 90.46 $\mu C/cm^2$ was calculated for the tetragonal phase. These results points to the potential of BiAlO$_3$ as a ferroelectric material, particularly if we, once again, trace a parallel between BAO and BFO. It is an established fact that ferroelectric properties can be enhanced through strain; for instance, polarizations values as large as 150 $\mu C/cm^2$ have been measured in BFO thin films, \cite{Yun2004,Ricinschi2006} so one can reasonably expect BAO  to exhibit a similar behavior. So far, the few measured values of polarization reported for BAO are significantly lower than those theoretically predicted. Further work will therefore be required in future  to assess the real potential of BAO as technologically relevant ferroelectric material.

\section{Electronic properties and Rashba-Dresselhaus spin-splitting effects}\label{sec:elec}

\begin{figure}
\centering
\includegraphics[width=1\linewidth]{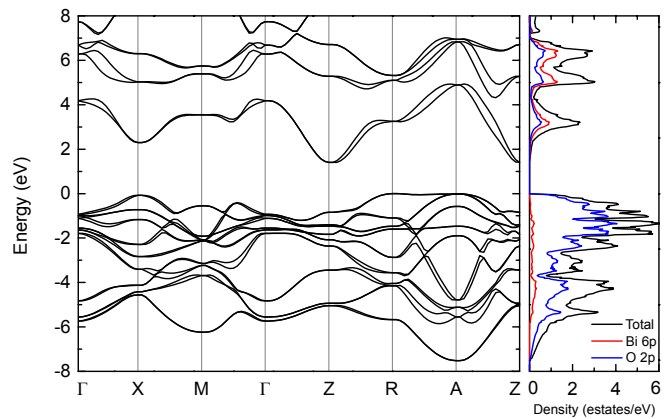}
\caption{Band structure, total density of states, and partial density of states for tetragonal BiAlO$_3$ calculated with GGA. Fermi level was set at the valence band maximum.}
\label{fig:P4mm-bands-dos}
\end{figure}

Figure \ref{fig:P4mm-bands-dos} shows the calculated band structure, total density of states, and partial density of states for the tetragonal phase. The upper valence band has a width of about 7.5 $eV$ and is mainly derived from O \textit{2p} orbitals. The maximum is located at the \textit{R} point, while the  \textit{A} and \textit{X} points are about 0.02 and 0.8 $eV$ below \textit{R}, respectively. The conduction band is made up mainly of Bi \textit{6p} orbitals with some O \textit{2p} admixture. The conduction band minimum is shifted from the \textit{Z} point by 0.026 \AA{}$^{-1}$ due to the Rashba effect. Moreover, the Rashba-like spin splitting is noticeable on the conduction band throughout the Brillouin zone, with the exception of the $\Gamma$$Z$ line, which is parallel to the polar axis [001] (consistently with the Rashba model where the spin-splitting occurs in the plane perpendicular to the electric field). The calculated indirect energy band gap is about 1.39 $eV$, somewhat lower than the ones reported in the literature for the hypothetical cubic phase. \cite{Wang200796,PhysRevB.75.245209} As PBE-GGA is known to underestimate the electronic band gap, the latter was also calculated by applying hybrid-functionals within the HSE approach. \cite{HSE2003} As expected, the calculated value of the gap increased to 2.57 $eV$.

\begin{figure}
\centering
\includegraphics[width=1\linewidth]{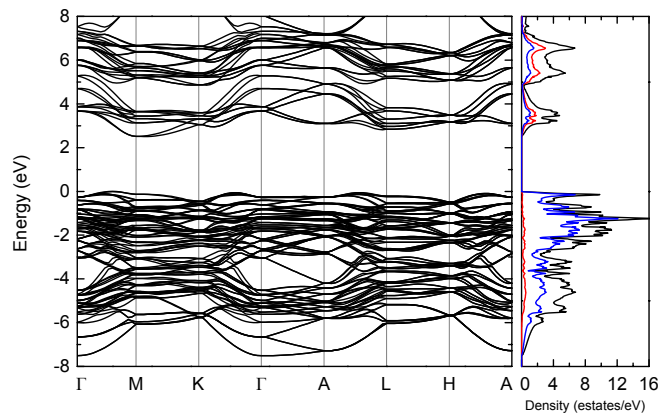}
\caption{Band structure, total density of states, and partial density of states for rhombohedral BiAlO$_3$ calculated with GGA. The bandstructure is presented using the hexagonal setting. Fermi level was set at the valence band maximum.}
\label{fig:R3C-bands-dos}
\end{figure}

The band structure and density of states for the rhombohedral phase , shown in Fig. \ref{fig:R3C-bands-dos}, are overall similar to those of the tetragonal phase, as far as the bandwidth and the orbital character of the valence and conduction bands are concerned. The valence band maximum is located between the points $\Gamma$ and \textit{M}, while the conduction band minimum is shifted from \textit{M} by 0.038 \AA{}$^{-1}$. The resulting indirect band gap of 2.52 $eV$ is significantly lower than the value of 3.28 $eV$ obtained (without SOC) in a previous study; \cite{Li2008539} this shows the importance of including SOC in the calculations of band structure of compounds with heavy elements, such as Bi. As in the tetragonal phase, a spin splitting is particularly noticeable around the conduction band minimum.

\begin{figure}
\centering
\includegraphics[width=1\linewidth]{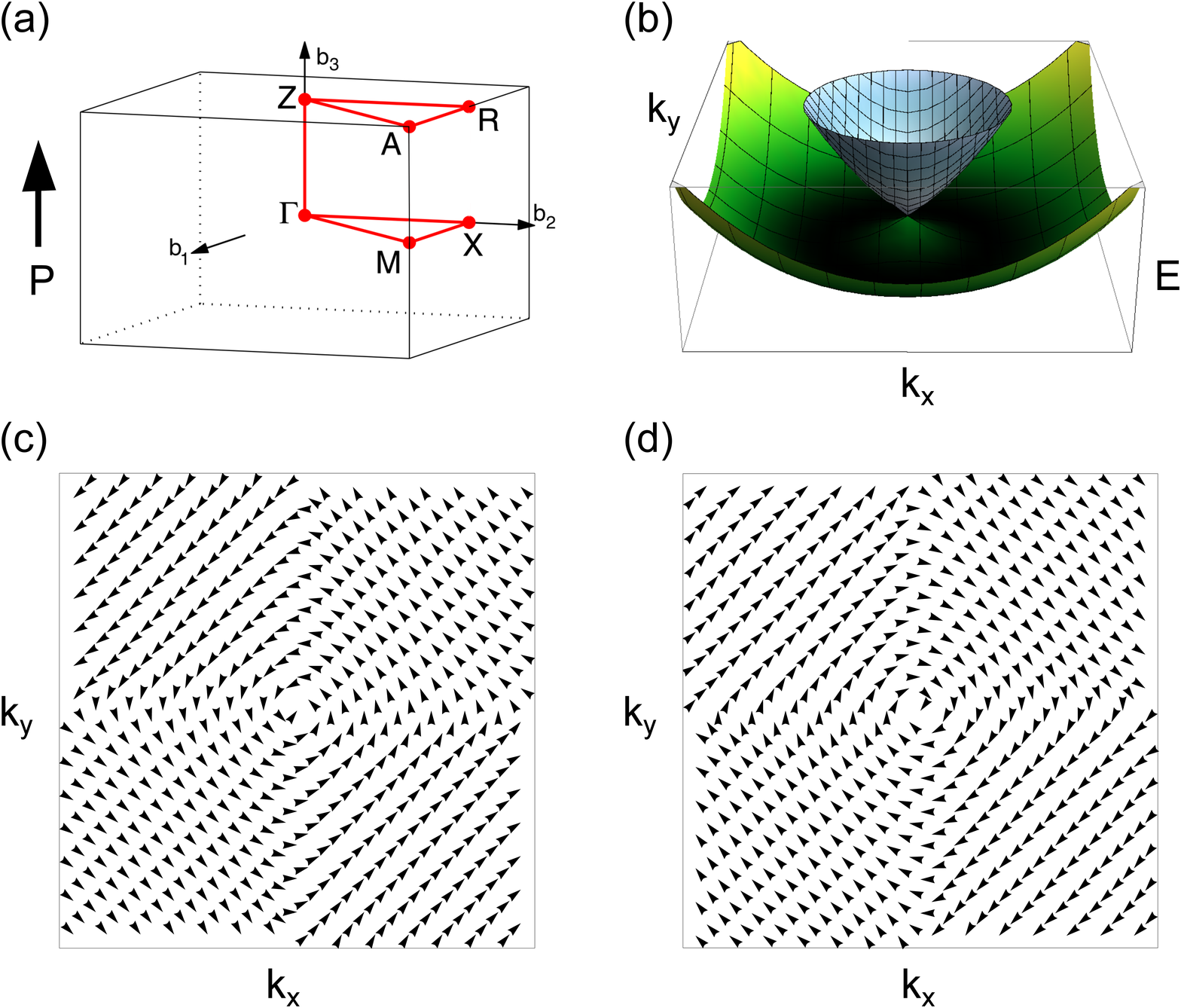}
\caption{(a) First Brillouin zone of the tetragonal structure (adapted from Ref. \onlinecite{Setyawan2010299}). Red lines highlight the path along which the band structure calculation was performed. P shows the direction of ferroelectric polarization. (b) Three dimensional energy dispersion close to the conduction band minimum: inner (outer) branch shown in light blue (green). (c) Spin texture of the outer branch of the conduction band minimum. (d) Spin texture of the inner branch of the conduction band minimum. Spin texture is calculated by computing spin expectation values on dense k-point mesh centered around Z, where the conduction band minimum is located.}
\label{fig:P4MM_textura}
\end{figure}

In order to better understand the nature of the observed spin splitting and spin-momentum coupling, we show in Fig. \ref{fig:P4MM_textura}(b)  the three-dimensional band structure for the tetragonal phase calculated around the \textit{Z} point, as well as the corresponding spin orientations (Figs. \ref{fig:P4MM_textura}(c) and \ref{fig:P4MM_textura}(d)). Here $k_x$ and $k_y$ are parallel to the reciprocal lattice vectors $b_1$ and $b_2$, respectively. Near the conduction band minimum, the bands are very similar to the parabolic energy dispersion of a 2DEG in a structure inversion asymmetric environment, characteristics of the $k$-linear Rashba effect, even though a small anisotropy is observed when looking at the position of the band minima along \textit{ZR} and \textit{AZ}. The spin textures (Figs. \ref{fig:P4MM_textura}(c) and \ref{fig:P4MM_textura}(d)) are also characteristic of a pure Rashba spin splitting, where the spin is always perpendicular to momentum. Moreover, no measurable out-of-plane spin component is observed. For deeper insights, we recall that the band dispersion around $Z$ in a plane orthogonal to the polar axis can be deduced by identifying all symmetry-allowed terms such that $O^\dagger H(k)O=H(k)$, where $O$ denotes all symmetry operations belonging to the little group of $Z$, supplemented by time-reversal symmetry. The little group of $Z$ $k$-point is $C_{4v}$, comprising two- and four fold rotation operations $C_2$, $C_4$ about the polar axis, two vertical mirror planes $M_1$, $M_2$ containing the principal axes, and two dihedral mirror planes $M_{d1}$, $M_{d2}$ containing the polar axis and bisecting the angle between $k_x$ and $\mp k_y$. Taking into account the transformation rules listed in Table \ref{tab:op}, the symmetry-allowed linear spin-momentum coupling has indeed the typical Rashba expression given in Eq. (\ref{H_Rashba}). 

As mentioned before, the strength of the Rashba effect can be quantified through several parameters, using the expression for the split bands in the nearly-free-electron approximation, Eq. (\ref{E_split}): the $k$-space shift $k_R$, the Rashba energy $E_R$, and the Rashba parameter $\alpha_R$. The parameters are estimated for Rashba-splitted bands along the direction in $k$-space in which the spin-splittings is larger, i.e., \textit{ZR} and \textit{AZ} for the tetragonal phase. $k_R$ is evaluated as the Rashba-induced momentum offset of the conduction band minimum with respect to the high-symmetry point \textit{Z}, while $E_R$ is calculated as the difference between the conduction band minimum estimated at $k_R$ and the corresponding energy values at the high-symmetry point. 
For the tetragonal phase we calculated $k_R=0.026$ \AA{}$^{-1}$ along the \textit{ZR} line, $E_R$ is estimated as 9.40 $meV$, leading to $\alpha_R=0.74$ $eV$\AA{}. Along \textit{AZ}, a similar value of 0.027 \AA{}$^{-1}$ was obtained for $k_R$, while the calculated smaller $E_R=$ 8.62 $meV$ would result in $\alpha_R=0.65$ $eV$\AA{}.
Such small but measurable anisotropy of the band splitting can be understood by considering higher-order terms in the effective model around point $Z$. The lowest higher-order term, which is compatible with the $C_{4v}$ symmetry, has the form $k_x^2k_y^2$, leading to
\begin{eqnarray}
H_Z(k) = E_0(k)+\mu_4\, k_x^2k_y^2 +\alpha(k)(\sigma_x k_y - \sigma_y k_x)
\end{eqnarray}
where $E_0(k)=\hbar^2(k_x^2+k_y^2)/2m^*$, the fourth-order term is parametrized by $\mu_4$, and the Rashba coupling constant $\alpha(k)=\alpha(1+\alpha_4\, k_x^2k_y^2)$ also contains a fourth-order correction. Clearly, the higher-order terms vanish along the $ZR$ ($k_y$) line, while they are expected to shift the energy position of the split band minimum along the diagonal $ZA$ line, meanwhile keeping the pure Rashba spin texture. Strictly speaking, the linear Rashba coupling constant  $\alpha$ corresponds to the Rashba parameter $\alpha_R$ evaluated along the $ZR$ line.

\begin{table}[b]
\caption{Transformation rules for crystal momentum $\boldsymbol{k}$ and spin-$1/2$ operators under the considered point-group symmetry operations. Time-reversal symmetry, implying a reversal of both spin and momentum, is defined as $i\sigma_yK$,  $K$ being complex conjugation and $\boldsymbol{\sigma}$ denoting Pauli matrices, while the point-group operations are defined as $C_2=e^{-i\sigma_z\frac{\pi}{2}}$, $C_4=e^{-i\sigma_z\frac{\pi}{4}}$, $M_1=i\sigma_y$, $M_2=i\sigma_x$, $M_{d1}=i(\sigma_x+\sigma_y)/\sqrt{2}$, and $M_{d2}=i(-\sigma_x+\sigma_y)/\sqrt{2}$. }\label{tab:op}
\begin{tabular}{lp{0.5cm}cp{0.3cm}c}
\hline
&&$\{k_x,k_y\}$ &&$\{\sigma_x,\sigma_y,\sigma_z\}$ \\
\hline
$C_2$ & &$\{-k_x,-k_y\}$ && $\{-\sigma_x,-\sigma_y,\sigma_z\}$\\
 $C_4$ & &$\{k_y,-k_x\}$& &$\{\sigma_y,-\sigma_x,\sigma_z\}$\\
  $M_1$&& $\{k_x,-k_y\}$ &&$\{-\sigma_x,\sigma_y,-\sigma_z\}$\\ 
  $M_2$ && $\{-k_x,k_y\}$ && $\{\sigma_x,-\sigma_y,-\sigma_z\}$\\
  $M_{d1}$ && $\{ -k_y,-k_x \}$ && $\{\sigma_y,\sigma_x,-\sigma_z\}$\\
  $M_{d2}$ && $\{ k_y,k_x \}$ && $\{-\sigma_y,-\sigma_x,-\sigma_z\}$\\
  \hline
\end{tabular}
\end{table}

\begin{figure}
\centering
\includegraphics[width=1\linewidth]{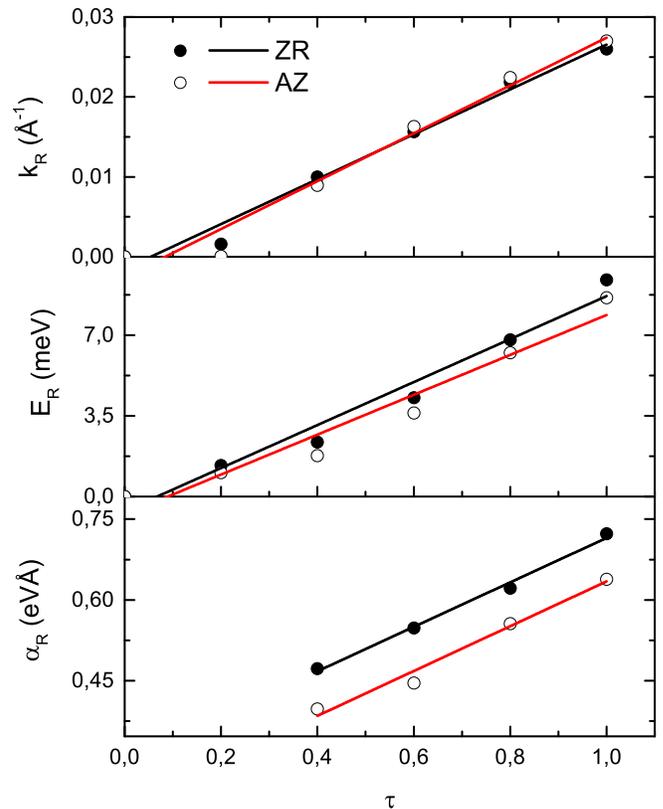}
\caption{Rashba parameters (momentum offset
$k_R$, energy splitting $E_R$, and Rashba parameter $\alpha_R=2E_R/k_R$) as a function of the ferroelectric order parameter $\tau$. Values calculated from the spin-splitting of tetragonal phase conduction band minimum along the symmetry lines ZR and AZ. Lines are guides to eyes.}
\label{fig:tau}
\end{figure}

The dependence of the Rashba parameters on the ferroelectric order parameter $\tau$ for the tetragonal phase was also analyzed ($\tau$ expresses the relative displacement between atoms in the unit cell). The results are shown in Fig. \ref{fig:tau}, in which a linear trend is observable for the splitting along both \textit{ZR} and \textit{AZ} symmetry lines. By tuning the $\tau$ parameter, we verified that the amplitude of the Rashba effect can be modulated accordingly. In particular, the spin degeneracy can be restored when the inversion symmetry is artificially brought back in the parent centrosymmetric structure. Importantly, our results also show that the Rashba parameter changes sign when the ferroelectric polarization is switched, implying a complete reversal of the spin-orientation texture (not shown), putting forward BAO as the first oxide candidate of the FERSC class of materials.

\begin{figure}
\centering
\includegraphics[width=1\linewidth]{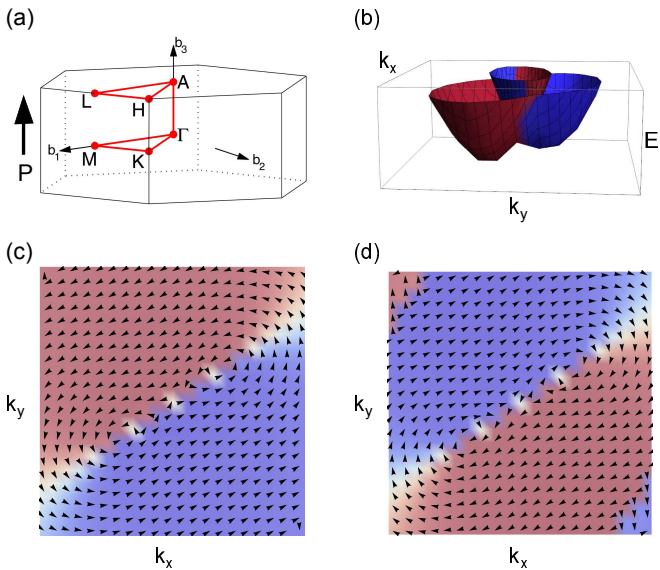}
\caption{(a) First Brillouin zone of the hexagonal structure (adapted from Ref. \onlinecite{Setyawan2010299}). Red lines highlight the path along which the band structure calculation was performed. P shows the direction of ferroelectric polarization. (b) Three dimensional energy dispersion close to the conduction band minimum. (c) Spin texture of the outer branch of the conduction band minimum. (d) Spin texture of the inner branch of the conduction band minimum. Red and blue colors indicate opposite out-of-plane spin components. Spin texture is calculated by computing spin expectation values on dense k-point mesh centered around M, where the conduction band minimum is located.}
\label{fig:R3C-texture}
\end{figure}

The rhombohedral phase exhibits more complex features. The band structure and spin textures calculated around the \textit{M} point are shown in Fig. \ref{fig:R3C-texture}. The spins possess a measurable out-of-plane component, which appears to be reversed when crossing the $\Gamma M$ line. The band structure has two minima with opposite out-of-plane spin components. The spin in-plane components, however, present the same vorticity. Both the strong anisotropy of the three-dimensional structure of the conduction bands around the $M$ point (Fig. \ref{fig:R3C-texture} (b)) and the non-trivial pattern of spin orientations (Figs. \ref{fig:R3C-texture} (c) and \ref{fig:R3C-texture}(d)) suggest that the spin-momentum coupling cannot be described by the simple expression of Rashba or Dresselhaus SOC given in Eqs. (\ref{H_Rashba}) and (\ref{H_Dresselhaus}). The effective low-energy Hamiltonian can be again deduced by considering the symmetry properties of the electronic wave vector around the high-symmetry point $M$. The little group of \textit{M} $k$-point comprises only a mirror operation about a plane containing the polar axis and the $\Gamma M$ line. For spin-$1/2$ electrons and taking $k_x$ parallel to the $\Gamma M$ line, the transformation rule for $M_1$ operation listed in Table \ref{tab:op} applies and the $\bm k\cdot\bm p$ Hamiltonian around $M$ assumes the following expression, including corrections up to linear order in $\bm k$:
\begin{equation}\label{H_r3m}
H_M(k)=E_0(k)+\alpha k_x\sigma_y + \beta k_y\sigma_x + \gamma k_y\sigma_z,
\end{equation}
where $\alpha$, $\beta$, $\gamma$ are three independent coefficients and the nearly-free-electron energy is $E_0(k)=\hbar^2(k_x^2/m^*_x+k_y^2/m^*_y)/2$. Notice that if $\alpha$=$-\beta$ and $\gamma$=$0$ one recovers the usual Rashba Hamiltonian Eq. (\ref{H_Rashba}), while for $\alpha$=$\beta$ and $\gamma$=$0$ the Hamiltonian describes the standard linear Dresselhaus coupling, as it transforms to the usual expression given in Eq. (\ref{H_Dresselhaus}) in a reference frame rotated by 45\textdegree. Therefore, Eq. (\ref{H_r3m}) describes a combination of Rashba-like and Dresselhaus-like linear spin-momentum couplings, with coupling constants given, respectively, by $(\beta-\alpha)/2$ and $(\alpha+\beta)/2$,  plus an additional anisotropic term with coupling constant $\gamma$; a similar form of the SOC Hamiltonian has been found to apply to electrons confined in (110) layers with zinc blende structure.\cite{cartoixa, gambardella_rev}
Solving the eigenvalues problem gives the following split energies:
\begin{equation}\label{eq:k-p}
E_\pm=E_0(k)\pm\sqrt{\alpha^2k_x^2+(\beta^2+\gamma^2)k_y^2}\equiv E_0(k)\pm E_s,
\end{equation}
corresponding to asymmetric dispersion relations which look like two partially overlapping parabolic cones, as in Fig. \ref{fig:R3C-texture}(b).
Interestingly, the additional term $\gamma k_y\sigma_z$ gives rise to a net momentum-dependent spin polarization along the $z$ axis. In fact, the averaged components of the spin operator $\langle\boldsymbol{\sigma}\rangle$ can be expressed as:
\begin{equation}\label{eq:sigma}
\begin{pmatrix}
\langle\sigma_x\rangle_\pm \\
\langle\sigma_y\rangle_\pm \\
\langle\sigma_z\rangle_\pm
\end{pmatrix}
=
\begin{pmatrix}
\pm\sin\theta\cos\xi \\
\pm\sin\theta\sin\xi \\
\pm\cos\theta
\end{pmatrix}
\end{equation}
where
\begin{equation}\label{eq:tan_theta}
\tan\theta = \frac{\sqrt{\alpha^2k_x^2+\beta^2k_y^2}}{\gamma k_y},
\end{equation}
\begin{equation}\label{eq:tan_xi}
\tan\xi=\frac{\alpha}{\beta}\cot\phi_k ,
\end{equation}
being $\boldsymbol{k}=|\boldsymbol{k}|(\cos\phi_k,\sin\phi_k)$. Specifically, the expectation value of spin-$z$ component reads $\langle\sigma_z\rangle_\pm=\pm\gamma k_y/E_s$, being reversed when moving across the mirror plane (while remaining opposite in outer or inner branches). On the other hand, the in-plane spin texture is mainly influenced by the relative sign of $\alpha$, $\beta$ coefficients, being more Rashba- or Dresselhaus-like if the ratio $\alpha/\beta$ is negative or positive, respectively. However, the in-plane components of the spin are expected to be rather small if $\gamma$ is much larger than $\alpha$, $\beta$.

By fitting the DFT energy dispersion along the $\Gamma M$ (i.e., $k_x$) and $MK$ (i.e., $k_y$) lines with Eq. \ref{eq:k-p}, one gets $\alpha$=$-0.082$ $eV$\AA{} and $\sqrt{\beta^2+\gamma^2}$=$0.408$ $eV$\AA{}. On the other hand, by combining Eqs. (\ref{eq:sigma}) and (\ref{eq:tan_theta}), the ratio between the coupling coefficients $\beta$, $\gamma$ can be estimated by considering the ratio between the in-plane and out-of-plane spin components along the line $MK$, where $k_x$=$0$, being
\begin{equation}
\frac{\beta}{\gamma}=\frac{\sqrt{\langle\sigma_x\rangle^2+\langle\sigma_y\rangle^2}}{\langle\sigma_z\rangle}=0.437.
\end{equation}
One finds, therefore, $\beta=0.163$ $eV$\AA{} and $\gamma=0.374$ $eV$\AA{}. We can compare these estimates for the linear spin-momentum coupling constants with the Rashba parameter $\alpha_R$ as evaluated from the ratio $2E_R/k_R$, where $E_R,\, k_R$ are calculated along the direction in $k$-space in which the spin-splitting is larger, i.e., the $MK$ line. We found $k_R$=$0.038$ \AA{}$^{-1}$, $E_R$=$7.34$ $meV$, resulting in a coefficient $\alpha_R$=$0.39$ $eV$\AA{}, to be compared with $\sqrt{\beta^2+\gamma^2}$=$0.408$ $eV$\AA{}. The fairly good agreement between the two estimates suggest that higher-order SOC terms are substantially ineffective in the rhombohedral phase. Eventually, we examined the spin-texture of the rhombohedral structure when switching the polarization direction. In analogy with the tetragonal phase, the spin-texture was fully reversed, pointing again to the possible control of the spin-texture by means of an external electric field.

\begin{table}[t]
\caption{Rashba parameters (momentum offset
$k_R$, energy splitting $E_R$, and Rashba parameter $\alpha_R=2E_R/k_R$) for BiAlO$_3$ in both tetragonal an rhombohedral phases, compared with corresponding data for selected systems.}\label{tab:rashba}
\begin{tabular}{p{2.4cm}cccc}
\hline
System & $k_R($\AA{}$^{-1})$ & $E_R (meV)$ & $\alpha_R (eV$\AA{}$)$ & Reference \\ 
\hline
BiAlO$_3$($R3c$) & 0.04 & 7.34 & 0.39 & This work \\
BiAlO$_3$($P4mm$) & 0.03 & 9.40$^\dagger$ & 0.74$^\dagger$ & This work \\
 & 0.03 & 8.62$^\ddagger$ & 0.65$^\ddagger$ & This work \\[0.1cm]
{\it Surfaces} &  &  &  &  \\
Au(111) & 0.012 & 2.1 & 0.33 & \cite{PhysRevLett.77.3419} \\
Bi(111) & 0.05 & 14 & 0.55 & \cite{PhysRevLett.93.046403} \\[0.1cm]
{\it Interfaces} & & & & \\
InGaAs/InAlAs & 0.028 & $<$1 & 0.07 & \cite{PhysRevLett.78.1335} \\[0.1cm]
{\it Bulk} & & & & \\
BiTeI & 0.052 & 100 & 3.8 & \cite{Ishizaka2011} \\
BiTeCl & $\sim$0.03 & 18.45 & 1.2 & \cite{xiang_prb2015}\\
BiTeBr & $<$0.05 & $<$50 & $<$2 & \cite{PhysRevLett.110.107204,Eremeev2013} \\
GeTe & 0.09 & 227 & 4.8 & \cite{Domenico2013} \\
SnTe & 0.08 & 272 &  6.8 & \cite{Evgeny2014}\\
$\beta-$(MA)PbI$_3$ & 0.015 & 12 & 1.5&\cite{Kim_pnas2014}\\
$\beta-$(MA)SnI$_3$ & 0.011 & 11 & 1.9 & \cite{Kim_pnas2014}\\
(FA)SnI$_3$ & 0.022 & 14.8 & 1.34 & \cite{Alessandro_ncomm2014}\\
LiZnSb 	& 0.023 & 21 & 1.82 & \cite{narayan2015}\\
KMgSb 	& 0.024 & 10 & 0.83 & \cite{narayan2015}\\
NaZnSb (PBE) 	& 0.024 & 31 & 2.58 & \cite{narayan2015}\\
NaZnSb (HSE)	 & 0.038 & 42 & 1.1 & \cite{Domenico2016}\\
\hline
{\footnotesize $\dagger$: along ZR}&&&&\\[-0.1cm]
{\footnotesize $\ddagger$: along AZ.}    
\end{tabular}
\end{table}

Before concluding, we summarize the Rashba parameters obtained in the present work for both phases  in table \ref{tab:rashba}. The parameters of a few selected bulk systems (mostly theoretically predicted in the framework of DFT calculations) are also shown for comparison. It can be seen that the Rashba parameters for BAO are comparable to the values reported for InGaAs/InAlAs interfaces (where SOC effects are expected to be relatively weak), and are much lower than those reported for other bulk systems (GeTe, Bismuth-telluro-halides, etc.) where a giant Rashba effect was invoked. This shows that  a large polarization (such as tens of $\mu C/cm^2$, as in BAO) and the presence of heavy elements (such as Bi in BAO) do not automatically imply a large Rashba spin splitting; rather, the orbital character of the involved electronic states, the related hybridizations allowed by the specific crystalline symmetry and/or local atomic structure, as well as the size of the gap, may play a relevant role.\cite{PhysRevB.84.041202}  This renders first-principles calculations - able to simultaneously and accurately describe all the above mentioned ingredients -  a useful tool for a careful estimate of the Rashba parameters.

\section{Conclusion}\label{sec:conclusion}
In summary, we have performed relativistic first-principles density functional calculations to systematically investigate the
electronic properties and the Rashba-like effect on the tetragonal and rhombohedral phases of the ferroelectric compound BiAlO$_3$. The ferroelectric polarization, band structure, densities of states were calculated, showing  - where available  - an overall good agreement with other studies on BAO reported in literature. Our results show that, near the Fermi level, the valence band is formed mainly by O \textit{2p} orbitals, while the conduction band is mainly derived from Bi \textit{6p} orbital with some admixture of O \textit{2p}. Fairly large indirect band gaps have been calculated for both phases, namely 1.49 $eV$ and 2.57 $eV$ for tetragonal and rhombohedral phases respectively. Our theoretical analysis indicates that non-negligible spin-splitting effects appear mostly near the conduction band minimum; furthermore, we argue that a bulk purely Rashba effect is responsible for calculated spin-splitting in the BAO tetragonal phase, while a sizable interplay of Rashba and linear Dresselhaus effects is observed in the BAO rhombohedral phase. In the latter case, the specific form of the spin-momentum interaction causes the spins to be oriented mainly parallel the polar axis. The full reversal of the spin texture with ferroelectric polarization switching was also predicted. BAO can therefore be considered as the first oxide showing a coexistence of ferroelectricity and Rashba-Dresselhaus effects. As previously noted, the prototypical FERSC, provided with an experimental confirmation of previous theory predictions, GeTe, has several drawbacks, mainly due to the large concentration of defects and related large conductivity, which often hinders the ferroelectric switching process. These disadvantages are not present in BAO, which is a well-known not-leaky ferroelectric for which the FE polarization switching has been experimentally proved. \cite{Son2008} Also, its band gap is much larger than the gap in GeTe (which is of the order of half an $eV$), preventing the difficulties related to leakage currents. Furthermore, BAO could be used as the insulating barrier in a tunnel junction, where a tunneling spin-Hall effect has been recently predicted to occur as arising from the noncentrosymmetric character of the barrier.\cite{Fabian_PRL2015} The possibility to tune the strength and the sign of the SOC coupling constants, and hence the tunneling spin-Hall currents, might offer appealing perspectives for novel spintronic devices.\cite{Marjana}However, BAO presents its own disadvantages. For example, the Rashba effect in BAO is much weaker than in GeTe, and the fact that the spin splitting is predicted to occur on the conduction band might require doping, in order to make its experimental detection more easily  accessible  (for example, via Angle-Resolved Photo-Emission Spectroscopy).

\section*{Acknowledgments}
We are grateful to Dr. Domenico Di Sante, Dr. Emilie Bruyer, Dr. Alessandro Stroppa, and Mrs. Danila Amoroso for many useful discussions. Support from CINECA Supercomputing Center (Bo, Italy), from MIUR under the PRIN project ``OXIDES'' is gratefully acknowledged. LGDS also thanks the Brazilian research funding agency CNPq for financial support within the ``Science Without Borders'' program, while PB acknowledges partial support from the European Union's Horizon 2020 research and innovation programme under grant agreement No 696656 GrapheneCore1.

\end{document}